\documentclass[apj]{emulateapj}
\usepackage{natbib}
\usepackage{mathptmx}
\usepackage{url}
\usepackage{graphicx}	% Including figure files
\usepackage{amsmath}	% Advanced maths commands
\usepackage{amssymb}
\usepackage[toc,page,titletoc]{appendix}
\usepackage{lineno}
\usepackage{longtable} 
\usepackage{booktabs}

\begin{document}
\nocite{*}
\title{\textbf{Formation of the dark-matter deficient S0 galaxy NGC 4111 under the tidal interactions}} %
%Article title
\author{
Mei Ai$^{1,2},$
Ming Zhu$^{1,2}$,
Nai-ping Yu$^{1,2}$,
Jin-long Xu$^{1,2}$,\thanks{E-mail: xujl@bao.ac.cn (Xu)}
Xiao-lan Liu$^{1,2}$,
Yingjie Jing$^{1}$,
Qian Jiao$^{3}$,
Yao Liu$^{4,5}$,
Chuan-peng Zhang$^{1,2}$,
Jie Wang$^{1}$,
Peng Jiang$^{1,2}$
}

\affil{$^{1}$National Astronomical Observatories, Chinese Academy of Sciences, 20A Datun Road, Chaoyang District, Beijing, People's Republic of China}
\affil{$^{2}$Guizhou Radio Astronomical Observatory, Guizhou University, Guiyang 550000, People's Republic of China}
\affil{$^{3}$School of Electrical and Electronic Engineering, Wuhan Polytechnic University, Wuhan 430023, People's Republic of China}
\affil{$^{4}$School of physicals and electronics. Qiannan Normal University for Nationalities, Longshan Road, Tuyun 558000, People's Republic of China}
\affil{$^{5}$Qiannan Key Laboratory of Radio Astronomy, Guizhou Province, Duyun, 558000, People's Republic of China}

%\affil[$\ast$]{}
%\affil[$\dagger$]{}

%%\email{$^{*}$aimei@nao.cas.cn}
\email{$^{*}$xujl@bao.ac.cn}
\email{$^{*}$mz@nao.cas.cn}

% These dates will be filled out by the publisher
\date{Accepted XXX. Received YYY; in original form ZZZ}

% Don't change these lines
\label{firstpage}
%%\pagerange{\pageref{firstpage}--\pageref{lastpage}}

\begin{abstract}

We present the high-sensitivity and large-scale atomic hydrogen (HI) observations towards lenticular (S0) galaxy NGC 4111 using the Five-hundred-meter Aperture Spherical Radio Telescope (FAST). The column density map shows that NGC4111 and seven other different types of galaxies share a huge HI gas complex. The data also suggest that NGC 4111 is interacting with seven galaxies. Moreover, we identified a rotating gas disk associated with NGC 4111 from the HI complex. Still, the HI disk rotation direction has deviated from its stellar disk about 34.2$^{\circ}$, indicating that the NGC 4111 galaxy is undergoing a transition from a spiral galaxy to an S0 galaxy by the tidal interactions. The obtained dark matter-to-stellar mass ratio of NGC4111 is 3.1$\pm$0.7, which is lower than the average value of the Local Universe, implying that the interactions may strip its dark matter. Our results suggest that in a galaxy group environment, tidal interactions have a significant effect on galaxy features. 
%group system that share one huge HI gas complex. We find that the NGC 4111 galaxy has experienced intense interactions with neighboring galaxies, and we explore its impact on the properties of the galaxy NGC 4111. Under the influence of interactions the NGC 4111 galaxy is undergoing a transition from a spiral galaxy to an S0 galaxy. Meanwhile most of the HI gas has been stripped out of the galaxy during this process. Besides its HI disk rotation direction has deviated from the stellar disk about 30$^{\circ}$ indicated by our simulation. In the meantime the dark matter of NGC 4111 is also stripped by interactions. Our results suggest that in galaxy group environment tidal interactions have significant effect on galaxy features. 
%In addition to that, with the highest sensitivity by now, the HI component for NGC 4143 and its HI gas bridge and SDSS J120625.42+422607.2 are detected for the first time. 

%A  about $\sim$ 120 kpc between NGC 4143 and the NGC 4111 system is also detected for the first time. 

%We also find that the HI gas for NGC 4111 is destructed by interactions with its neighbour. HI gas center for NGC 4111 has deviated from its optical disk by 1.57'.The HI gas rotation disk differs from its optical disk about 40$^{\circ}$. Our results implies that galaxies in group environment had undergone multiple and complicated interactions. 

\end{abstract}

%%%%%%%%%%%%%%%%%%%%%%%%%%%%%%%%%%%%%%%%%%%%%%%%%%

%%%%%%%%%%%%%%%%% BODY OF PAPER %%%%%%%%%%%%%%%%%%

\section{Introduction}

%In the local Universe, lenticular (S0) galaxies 

How the lenticular (S0) galaxy formed is one of the key questions in galaxy evolution because S0s lie at the transient stage between spirals and ellipticals on the Hubble sequence. They exhibit a bulge in the central region and a disk without spiral arms. Multiple mechanisms are proposed to explain the formation of S0 galaxies (\citealt{Balogh2000}; \citealt{Just2010}; \citealt{Rizzo2018}; \citealt{Deeley2020}). It has long been known that there is an anticorrelation between spirals and S0s fractions over cosmic time in high-density cluster environments \citep{Dressler1980}. It is natural to speculate that the S0 galaxies originate from spirals. This is one of the main pathways for S0 formation, referred to as the ``faded spiral" mechanism (\citealt{Quilis2000}; \citealt{Crowl2005}; \citealt{Moran2007}; \citealt{Laurikainen2010}; \citealt{Cappellari2011}). In this picture, the gas of a spiral galaxy is stripped when it falls into a cluster environment, called ram pressure stripping \citep{GunnGott1972}. Most of the gas is stripped by the hot intracluster medium, and the future star formation is quenched. This process is relatively moderate, and the galaxy disk is not destructive \citep{Bekki2011}. Normally, the left S0 galaxy would retain the rotation feature of the predecessor spiral \citep{Deeley2021}. Another major pathway for the formation of the S0 galaxy is through mergers (\citealt{Spitzer1951}; \citealt{Diaz2018}; \citealt{Mendez-Abreu2018}). Both simulations and observations have shown that the S0 galaxy could be produced by major mergers(\citealt{Bekki1998}; \citealt{Eliche-Moral2018}). This kind of event often occurs in low-density environments and its angular momentum is decreased \citep{Querejeta2017}.  In addition to the two main pathways, the galaxy-galaxy interactions in the group environment are more effective in transforming spirals into S0s (\citealt{Just2010}; \citealt{Bekki2011}; \citealt{Xu2023}). Further exploration is needed to address the significant question of how spiral galaxies lose their arms and cease star formation. 

%other mechanisms might also be active. One of these mechanisms is that group environment also promotes the formation of S0 galaxies. Some observations have shown that S0 fractions grow more rapidly in small groups than rich clusters over cosmic time (\citealt{Just2010}; \citealt{Xu2023}), which implies that galaxy-galaxy interactions in the relatively low group environment are more effective in transforming spirals into S0s. Further exploration is needed to address the significant question of how spiral galaxies lose their arms and cease star formation. 

The NGC 4111 galaxy group system is the second most massive subgroup in the Ursa Major Cluster after M 109. According to \citep{Makarov2011}, there are 20 member galaxies in the NGC 4111 group. In this small group environment, galaxy-galaxy interactions are supposed to be more prominent than galaxy-environment interactions. The NGC 4111 galaxy is S0 type and is the largest galaxy in this system. This makes NGC 4111 an excellent case for studying environmental effects on the formation of S0. NGC 4111 has a polar ring at the center that is perpendicular to its main stellar disk. This ring probably originates from a minor merger with a gas-rich dwarf \citep{Roier2022}. The stellar age distribution analysis also suggests that NGC 4111 has undergone a minor merger \citep{Kasparova2016}. NGC 4111 is also surrounded by companion galaxies located within 40-200 kpc, which implies that tidal interactions would also be important. 

Neutral hydrogen (HI) gas that is loosely gravitationally bound to its host galaxy is one of the most powerful tools in studying galaxy interactions. The HI gas properties could provide us with information on the gas disk kinematics and rotation features. The diffuse HI tidal tails around galaxies would indicate the possible interaction history. The NGC 4111 system was observed by the Lovell telescope at Jodrell Bank Observatory (UK) in 2013 \citep{Wolfinger2013}. It was also the target of the Westerbork Synthesis Radio Telescope (WSRT, \citealt{Serra2012}), as part of the ATLAS$^{\rm 3D}$ project. Both results confirm the huge HI gas complex around NGC 4111 and the WSRT data resolve the high-density HI gas region. In this Letter, we use FAST which is the most sensitive telescope at present to re-observe the NGC 4111 group system. We expect that more diffuse HI gas detected would enable us to explore the interaction processes and investigate the impact on the formation of S0 galaxies.   

%The results show that there is a huge HI gas envelope that covers this group. While diffuse HI gas clouds in the outer regions are missed due to the sensitivity issue. 

%diffuse structures of the member galaxies are invisible due to the limitation of resolution.
%TPrevious HI observations for NGC 4111 system have shown that part of the member galaxies share one huge HI gas complex (Wolfinger et al. 2013). 

%The angular resolution of the Lovell telescope at 1.4 GHz is 12.0 arcmin and the rms noise level in the HIJASS cube is 12-14 mJy/beam (Lang et al. 2003, Wolfinger et al. 2013). Their results show that there is a huge HI gas envelope that covers this group. Detailed structures of the member galaxies are invisible due to the limitation of resolution. 
%The neutral hydrogen (HI) gas emission line is one of the most powerful tools in studying the galaxy accretion and interactions. They are loosely gravitationally bound to its host galaxy and easily disturbed (Hibbard van Gorkom 1996, Yun 1996). The properties of the galaxy HI gas contain the information that how the galaxy is interacted with the environment. With the improved sensitivity and resolution of modern radio telescopes, more and more faint HI gas are detected and help to shed light on the evolution process of galaxies.

%Figure 1
\begin{figure*}
\centering
\includegraphics[width=0.7\textwidth]{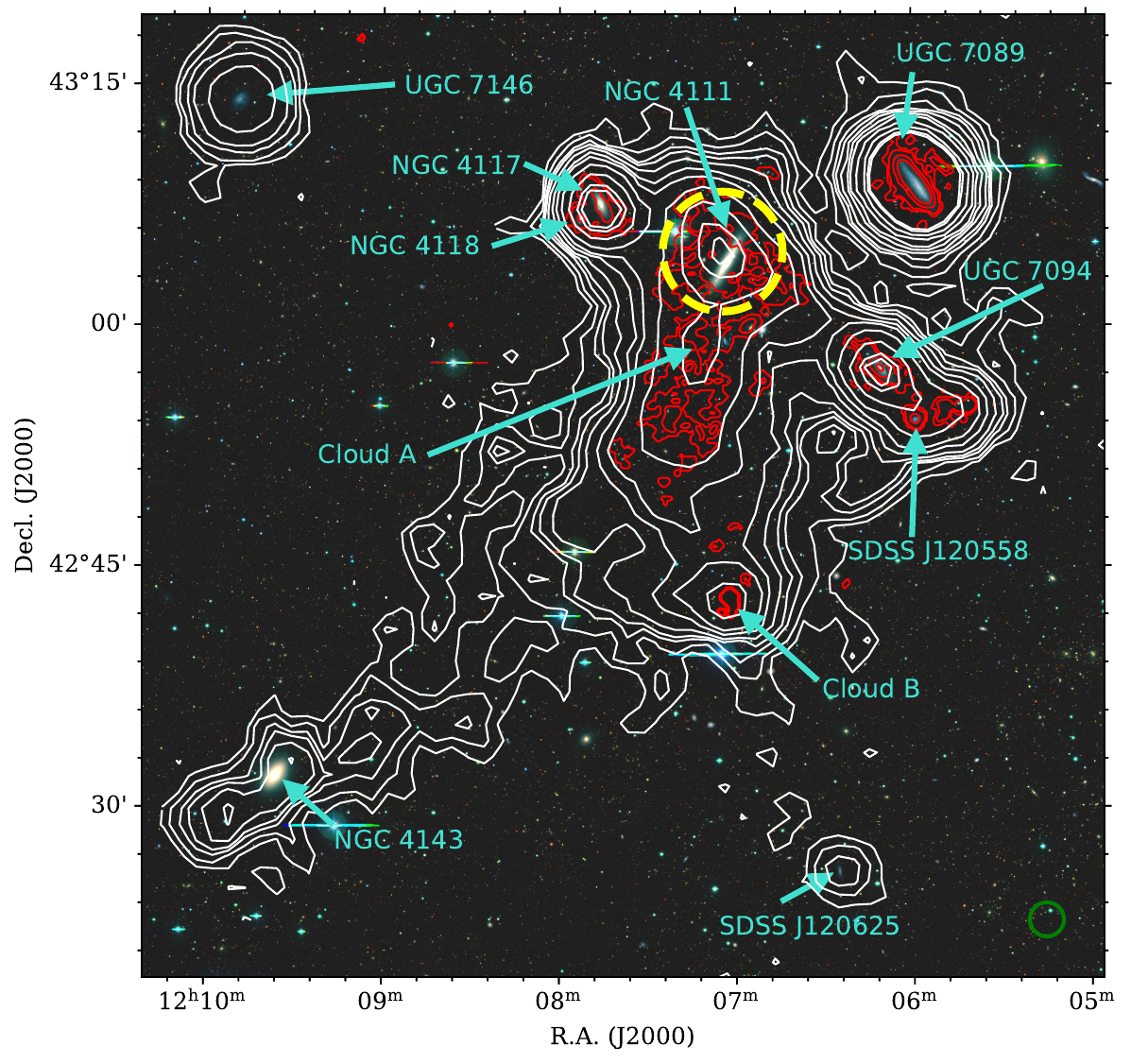}
\caption{The NGC 4111 galaxy group system. White contours show the HI column density map for this system derived from FAST data, overlaid on the DECaLs optical image. Nine galaxies are on this map and their names are labelled. The velocity range for the integration is from 600 to 1200 km $\rm s^{-1}$. The FAST beam size is labelled as a green circle at the bottom right corner. The HI contour levels are: 1.9$\times10^{18}\rm cm^{-2}$, 3.4$\times10^{18}\rm cm^{-2}$, 5.5$\times10^{18}\rm cm^{-2}$, 7.0$\times10^{18}\rm cm^{-2}$, 9.6$\times10^{18}\rm cm^{-2}$, 1.4$\times10^{19}\rm cm^{-2}$, 1.8$\times10^{19}\rm cm^{-2}$, 2.8$\times10^{19}\rm cm^{-2}$, 4.4$\times10^{19}\rm cm^{-2}$, 6.3$\times10^{20}\rm cm^{-2}$, 7.3$\times10^{19}\rm cm^{-2}$. The red contours are derived from the WSRT ATLAS$^{\rm 3D}$ public data, the column density levels are: 2.7$\times10^{19}\rm cm^{-2}$, 1.4$\times10^{20}\rm cm^{-2}$, 5.4$\times10^{20}\rm cm^{-2}$, 1.4$\times10^{21}\rm cm^{-2}$. }
\label{fig1}
\end{figure*}

\section{Observations and data reduction}

The data of the NGC 4111 group were first taken with the drift scan mode as part of the FAST All-Sky HI Survey (FASHI, \citealt{Zhang2023}), which is a new large sky survey for HI emission in the northern sky ($60^{\circ}$ $>$ Decl. $>$ $-10^{\circ}$) over the
velocity range -2,000 to 20000 km $\rm s^{-1}$. More subsequent observations are also performed for this source. The observations described here were taken in May and December 2021 and January and May 2022 with drift scan and On-the-Fly Map (OTF) mode. For all the observations we use the focal-plane 19-beam receiver which works in the frequency range from 1000 MHz to 1500 MHz with dual polarization mode. We utilize the backend of a wide band spectrometer (SpecW), which has 65536 channels and covers 500MHz bandwidth for each polarization and beam. The velocity resolution for this backend is 1.67 km $\rm s^{-1}$. The beamwidth for FAST is about 2.9$^{\prime}$ at 1.4GHz for each beam and the pointing accuracy is about 10$^{\prime\prime}$(\citealt{Jiang2019}; \citealt{Jiang2020}). The beam array was rotated by 23.4 degrees so that the beam tracks were equally spaced. The system temperature ranges between 18-22 K for both observational modes. For the observational strategy of HI, we choose to record data every 1s and inject a 10 K noise signal (CAL) every 32 s for calibration. 

The data were reduced by the HiFAST pipeline reduction software, which is developed by \citealt{Jing2024}. HiFAST automatically performed the flux calibration \citep{Liu2024}, baseline subtraction, RFI flagging, standing wave fitting \citep{Xu2024}, Doppler correction, and imaging procedure. Baseline correction was performed using the asymmetrically re-weighted penalized least squares smoothing algorithm (arPLS, \citealt{Baek2015}). In the re-gridding and imaging procedure, about 40-60 spectra would be assigned to one pixel in the non-edge cube area. The output from the HiFAST pipeline is a standard FITS data cube. The pixel size of the cube is set to $1'$ for an optimal exhibition of the data. The velocity range is cut between 200 to 2000 km $\rm s^{-1}$ for our data cube. To investigate the dark and diffuse source, the whole region is observed 6 times either by drift scan or OTF mode. Moreover, the data cube is Gaussian smoothed to 8.35 km $\rm s^{-1}$. After this, the final cube has the one sigma noise (rms) of about 0.5 mJy/beam or 1.73$\times10^{17}\rm cm^{-2}\rm channel^{-1}$.

%The sky region of the data cube spans from $12^{\rm h}05^{\rm m}$ + $42^{\circ}15'$ to $12^{\rm h}11^{m}$ + $44^{\circ}05'$ which covers the NGC 4111 complex HI gas and its neighbor NGC 4138. HIFAST provides different methods for fitting standing wave. Here we choose to use the Fast Fourier Transform (FFT-filter) method to remove the standing wave in our data. The FFT-filter method first preform Fourier transform to the raw spectra so the standing waves could be selected in the Fourier space. And then by conducting an inverse Fourier transformation we obtain the standing wave spectra selected before. More details about standing wave removal are referred to Xu et al. (Xu et al. in preparation). The RFI mitigation techniques are described in \citet{Zhang2021}. These spectra are weighted averaged by a Bessel$\times$Gaussian kernel to one pixel spectrum (Mangum et al. 2007).
%Some dark and small HI galaxies are also presented in this cube.

\section{Results}
%Figure 2
\begin{figure*}
    \centering    
    % 第一行左图
    \hspace{-1.25cm}
    \begin{minipage}[t]{0.4\textwidth} % [t] 确保顶部对齐
        \centering
        \includegraphics[width=9cm]{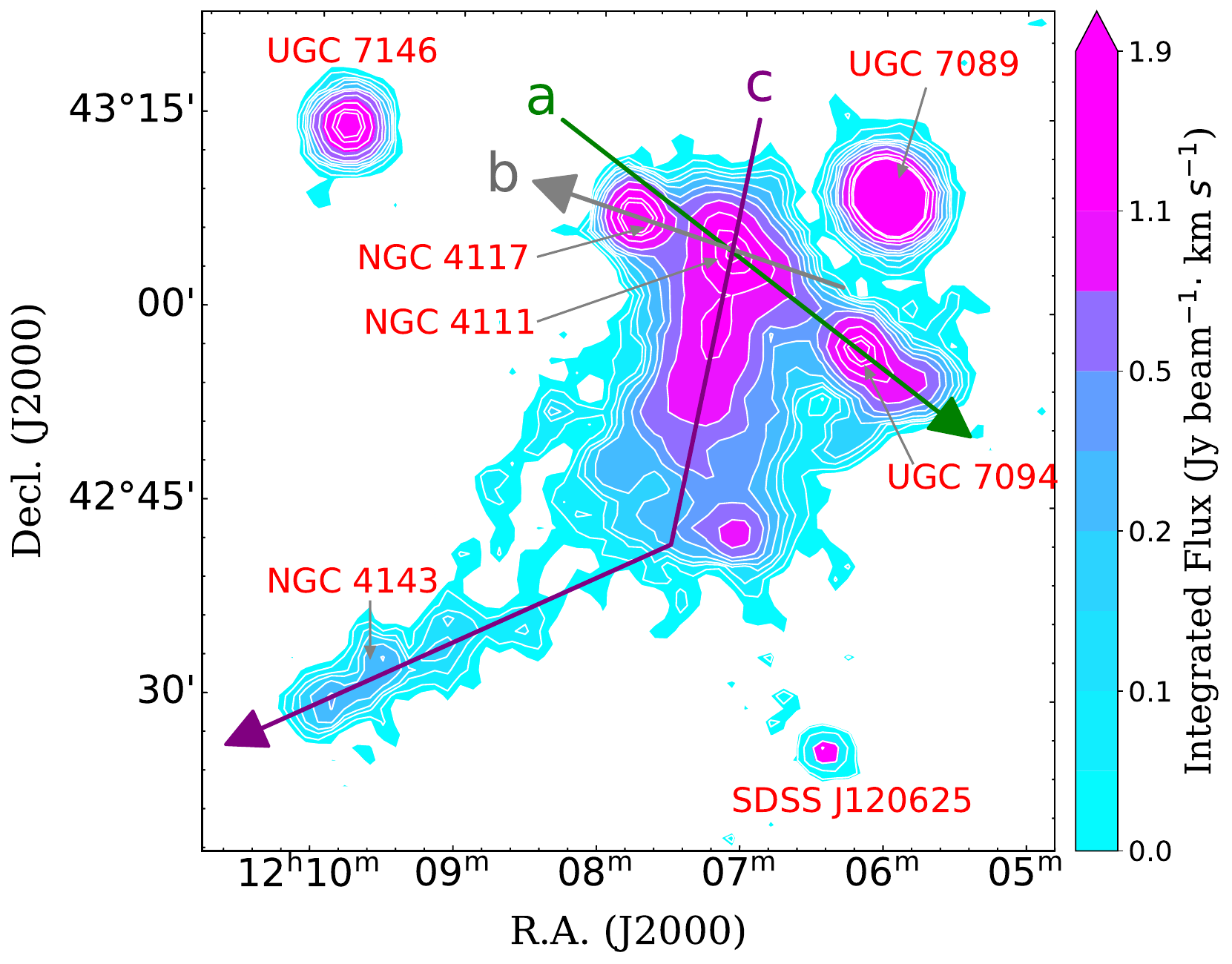} 
    \end{minipage}
    %\hfill % 自动调整水平间距
    % 第一行右图
     \hspace{1.8cm}
    \begin{minipage}[t]{0.4\textwidth} % [t] 确保顶部对齐
        \centering
        \includegraphics[width=9cm]{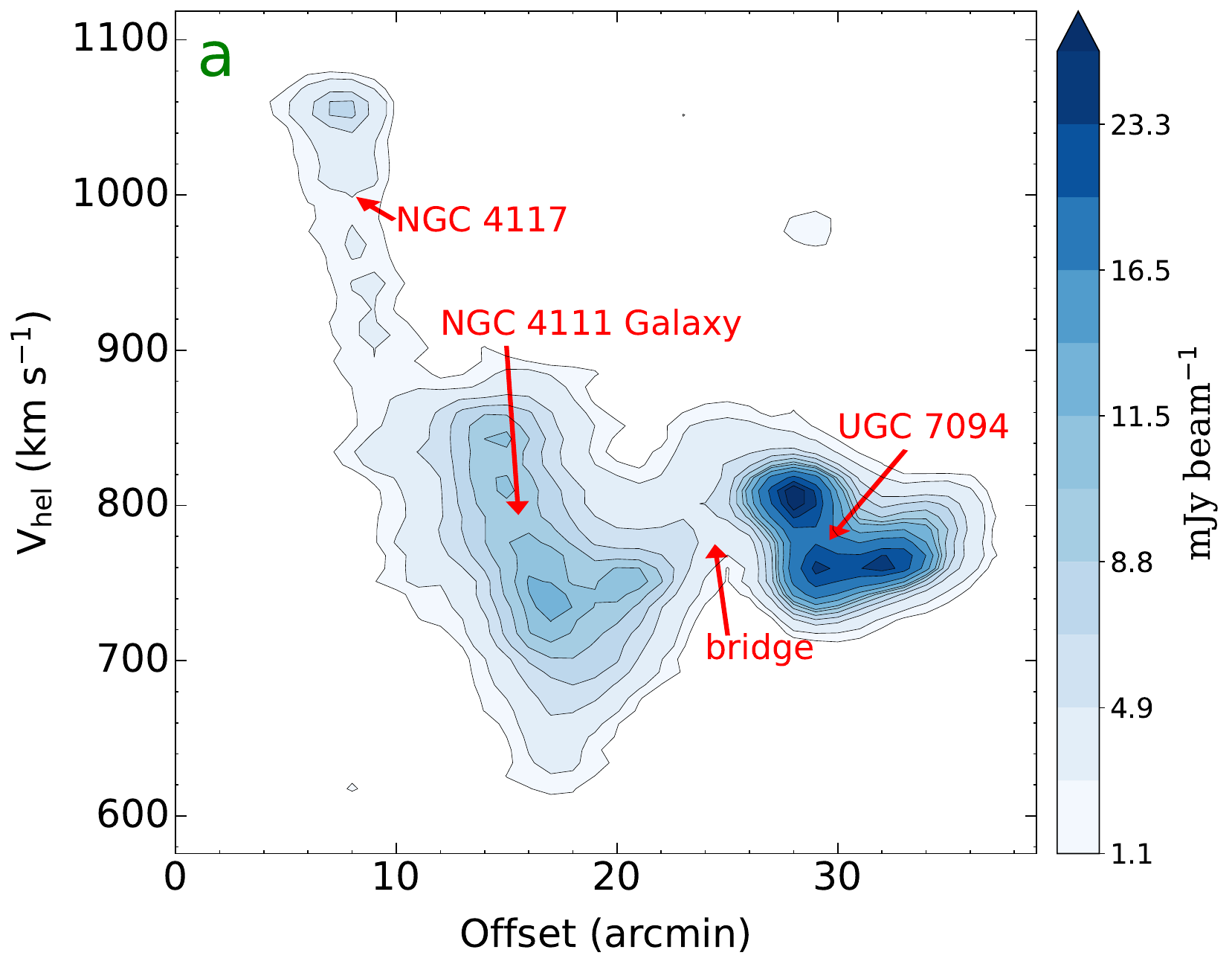}
    \end{minipage}   

    \vspace{0.5cm} % 调整第一行和第二行之间的垂直间距

    % 第二行左图
    \hspace{-1.25cm}
    \begin{minipage}[t]{0.4\textwidth} % [t] 确保顶部对齐
        \centering
        \includegraphics[width=9cm]{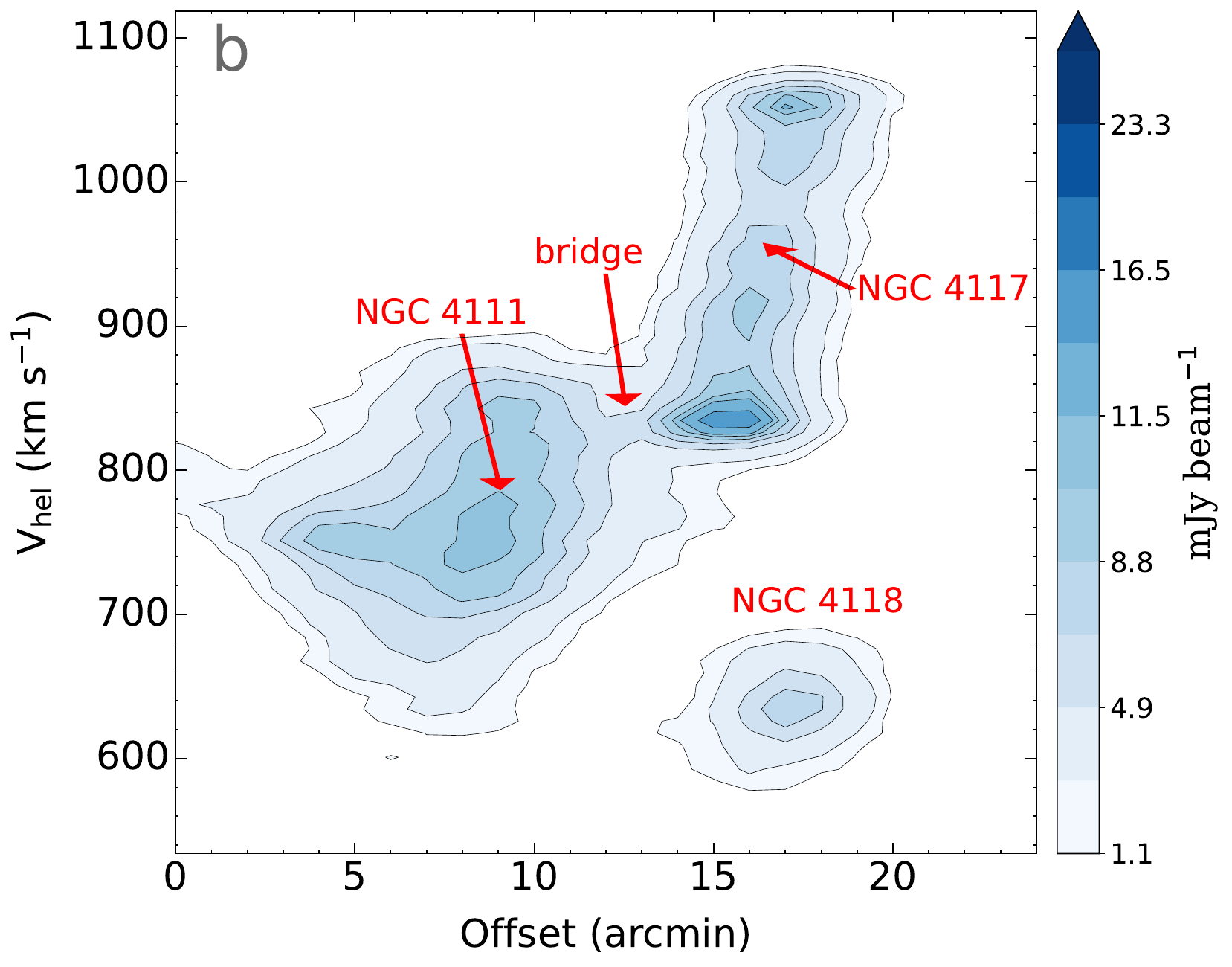}  
    \end{minipage}
    %\hfill % 自动调整水平间距
    % 第二行右图
    \hspace{1.8cm}
    \begin{minipage}[t]{0.4\textwidth} % [t] 确保顶部对齐
        \centering
        \includegraphics[width=9cm]{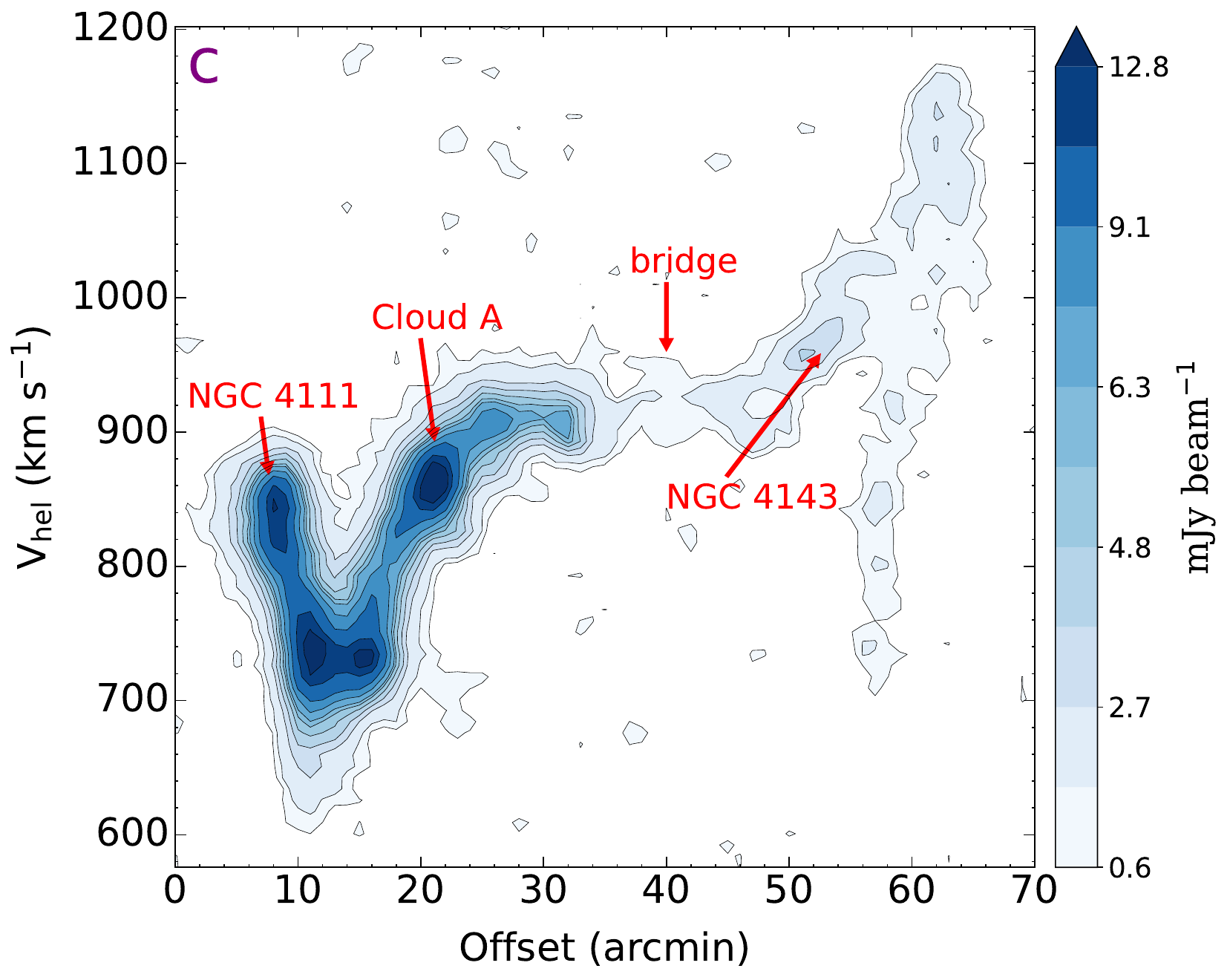}
    \end{minipage}
    \caption{ Three position-velocity (P-V) plots in three directions that centered on NGC 4111, illustrating its interactions with UGC 7094, NGC 4117, and Cloud A in radial direction. The upper left panel shows the HI moment 0 map for the NGC 4111 system. Three directions \textbf{a}, \textbf{b}, and \textbf{c} are overlaid around NGC 4111. The contour levels are the same as in Figure 1. The rest three panels exhibit the P-V diagrams along the directions \textbf{a}, \textbf{b}, and \textbf{c}. The contour levels for the three P-V diagrams are: 1.1, 2.2, 4.9, 6.9, 8.8, 10.4, 11.5, 16.4, 20.5, 23.29, 25.2 mJy beam$^{-1}$.}
\end{figure*}
%Figure 2

%Figure3
\begin{figure*}
    \centering
    % 左图
    \begin{minipage}[]{0.45\textwidth}
        \hspace{-1.0cm} % 左图水平移动
        \includegraphics[scale=0.38]{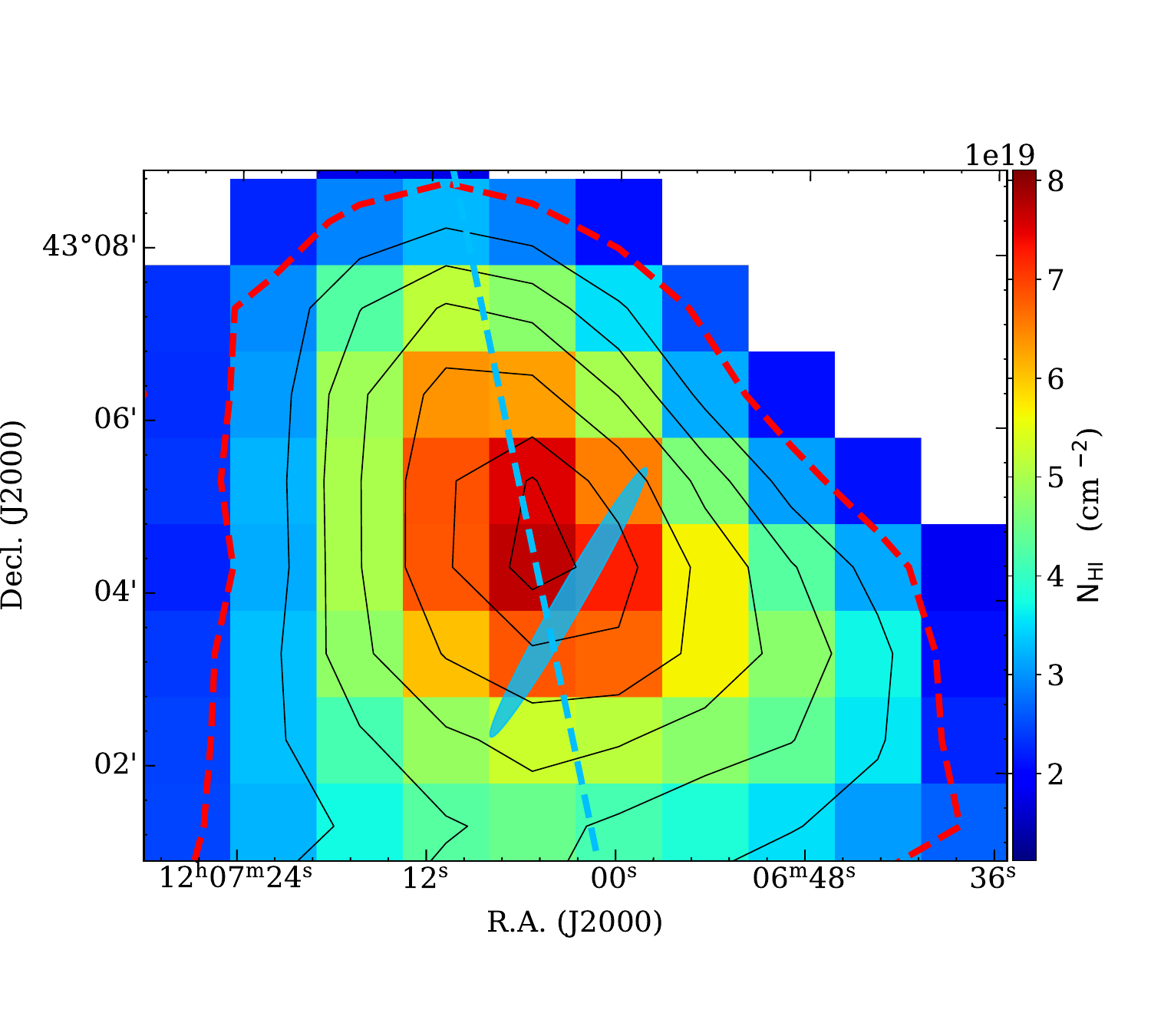}        
    \end{minipage}
    \hspace{0.6cm} % 左右图之间的水平间距
    % 右图
    \begin{minipage}[]{0.45\textwidth}       
        \vspace{0.9cm} % 向下移动右图
        \centering
        \includegraphics[width=\textwidth]{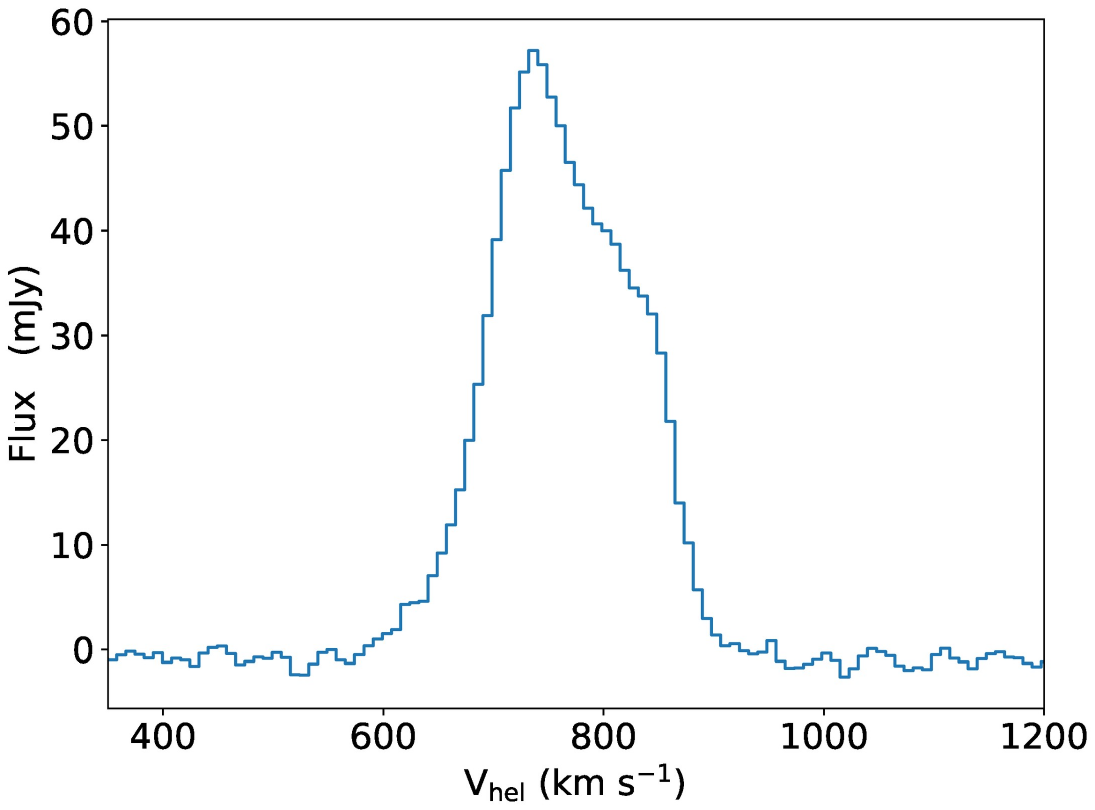}
    \end{minipage}    
    %\vspace{-120.8cm} % 整体垂直间距调整
    \setlength{\abovecaptionskip}{-15pt}
    \caption{ The method for extracting the HI integral spectrum for NGC 4111 from the entire HI complex. Left: The HI column density map around the NGC 4111 region integrated from 600 to 900 km $\rm s^{-1}$. The blue ellipse shape marks the optical disk of NGC 4111, the dashed blue line indicates the direction of the simulated major HI disk, see Section 4. The contour levels are: 2.5$\times10^{19}\rm cm^{-2}$, 3.3$\times10^{19}\rm cm^{-2}$, 4.1$\times10^{19}\rm cm^{-2}$, 4.9$\times10^{19}\rm cm^{-2}$, 5.8$\times10^{19}\rm cm^{-2}$, 6.8$\times10^{19}\rm cm^{-2}$, 7.4$\times10^{19}\rm cm^{-2}$; The right panel is the integral HI spectra of NGC 4111, which is the sum of all the spectra that inside of the lowest contour level (the red dashed contour line) in the left panel.}
\end{figure*}
%Figure3

\begin{table*}
    \caption{Major parameters of the member galaxies in the NGC 4111 system}
    \label{table 1}
    \setlength{\tabcolsep}{6.4pt}
    \begin{tabular}{lccccccccccccc} % four columns, alignment for each
        \hline
        Name &$d_{25}$&P.A.&incl&Type&$\textit{g} - \textit{r}$ & $\rm M_r$&$\rm M_*$&$V_{\rm sys}$&$w_{50}$&$M_{\rm HI}$&$f_{\rm HI}$\\ 
         & log(0.1arcmin)&degree&degree&&mag&mag&$\rm logM_{\odot}$&km/s&km/s&$\rm logM_{\odot}$&$\%$\\
         \hline
        NGC4111&1.25&151.4&90&S0&0.61&-19.70&10.11&778.5$\pm$0.5&180.0$\pm$1.1&8.6$\pm$0.06&2.3      \\
        UGC 7094&1.09&38.3&90&Sm&0.54&-15.69&8.43&780.3$\pm$0.3&75.6$\pm$0.5&8.5$\pm$0.06&50.1\\
        NGC 4117&1.20&20.5&90&S0&0.77&-19.08&10.04&929.8$\pm$6.6&256.7$\pm$9.6&8.27$\pm$0.2&4.0\\
        NGC 4118&0.83&148.3&64.4&dE&0.46&-15.20&8.15&635.0$\pm$3.8&60.8$\pm$11.3&7.6$\pm$0.1&17.9\\
        NGC 4143&1.42&144.0&68.3&S0&0.84&-20.56&10.71&975.8$\pm$7.0&297.7$\pm$37.2&8.0$\pm$0.08&0.4 \\
        UGC 7089&1.51&36.0&90&Sdm&0.46&-17.09&8.91&757.3$\pm$14.5&76.6$\pm$58.7&9.0$\pm$0.1&83.4\\
        UGC 7146&1.02&130.9&47.7&SABm&0.36&-16.26&8.46&1055.0$\pm$0.3&70.9$\pm$1.5&8.3$\pm$0.06&153.7\\
        SDSS J120625 &0.55&123.1&55.5&dI&0.22&-13.76&7.31&874.1$\pm$4.2&45.9$\pm$6.9&6.9$\pm$0.05&40.1\\
        SDSS J120558 &0.69&105.4&46.3&dI&0.32&-13.86&7.46&-&-&-&-\\
        \hline
    \end{tabular}
    \footnotetext[1]{ The diameter data $d_{25}$, the position angle and the inclination are obtained from the HyperLeda database. $d_{25}$ is in unit of $\log(0.1')$ and position angle is in North Eastwards direction.}
    \footnotetext[2]{ The morphological types data and the \textit{g} and \textit{r} band photometric data are from the SIMBAD database.}
    \footnotetext[3]{ $V_{\rm sys}$, $w_{\rm 50}$, $\rm M_{\rm HI}$ and $f_{\rm HI}$ for the member galaxies are measured in this work. See the text for details.}
    \footnotetext[4]{ The full galaxy names for SDSS J120625 and  SDSS J120558 are SDSSJ120625.42+422607.2 and SDSSJ120558.90+425406.2. It should be noted that they are from the SDSS DR7 database. The names have been updated to SDSS J120625.35+422604.8 and SDSS J120559.54+425409.3 in the DR18 database, respectively. }
\end{table*}

\begin{figure*}
\centering
    \includegraphics[width=0.45\textwidth,height=59mm]{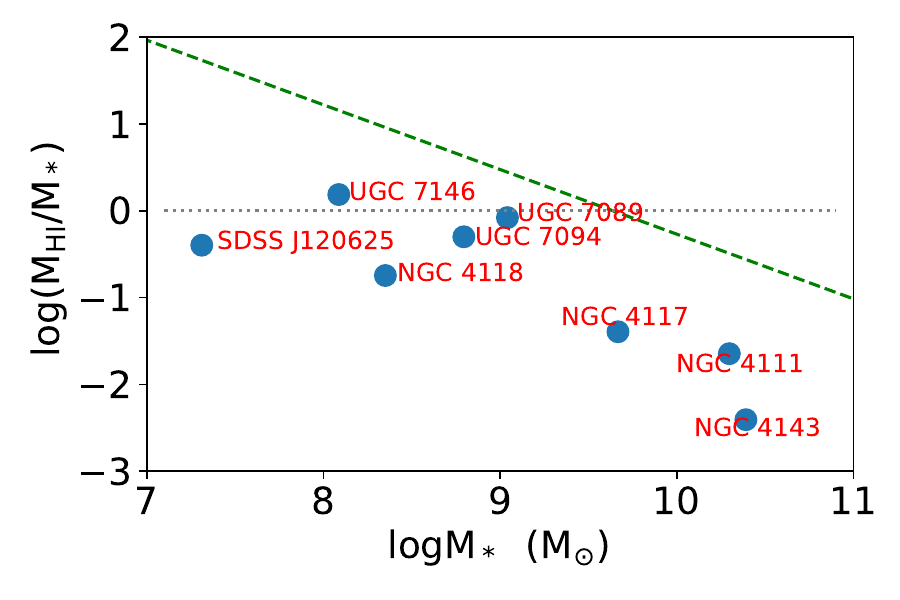}
    \includegraphics[width=0.45\textwidth,height=60mm]{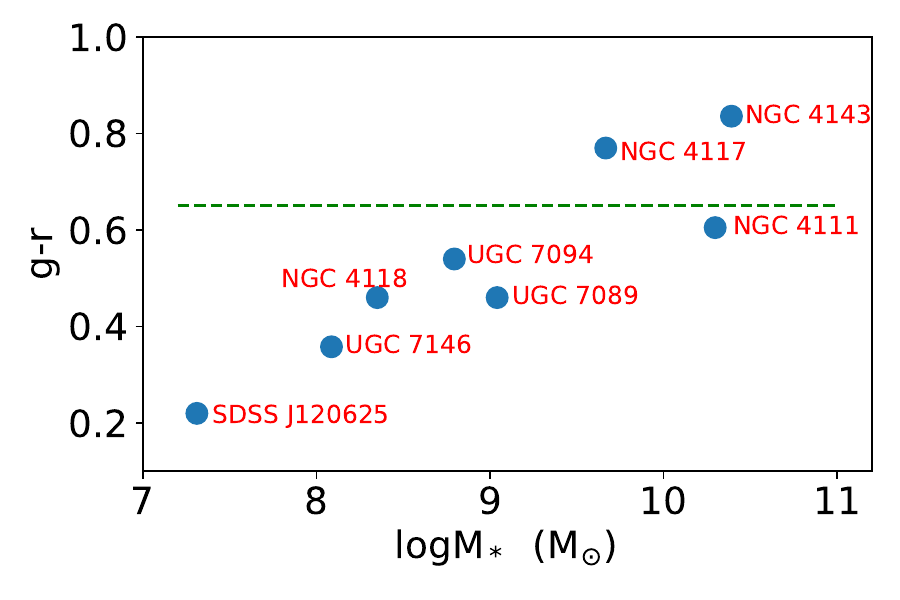}
\caption{ The scaling relations for the member galaxies in the NGC 4111 group system. Left: The correlation between HI mass fraction and stellar mass. The black dotted line marks the place where the galaxy stellar mass equals the HI gas mass \citep{Lelli2022}. The green dashed line indicates the typical correlation between HI mass fraction and stellar mass. Right: The correlation between stellar masses and \textit{g} -  \textit{r} colors for the member galaxies. The green dashed line is the constant \textit{g} -  \textit{r} value of 0.65, which we take to differentiate from blue to red galaxies \citep{Nelson2018}.}
\label{fig4}
\end{figure*}

In Figure 1 we show  the HI column-density map (moment 0 map) in white contours, overlaid on the optical DECaLS image \citep{Dey2019}. The white contours are derived from our FAST observations integrated from 600 to 1200 km $\rm s^{-1}$. The red contours are from the public WSRT ATLAS$^{\rm 3D}$ moment 0 data. The WSRT data resolve the high-density gas region, and FAST detects much more diffuse gas around this system. We detected HI emission in NGC 4143 and SDSS J120625.42+422607.2 for the first time. Eight galaxies share one huge HI envelope, which suggests that they are likely interacting together.

%The total mass of the NGC 4111 HI complex is 10$^{9.28} \rm M_{\odot}$ after subtracting the mass of UGC 7146 as shown in Figure 1.
%The tail in the west of UGC 7146 is a foreground intergalactic gas cloud. The angular length from UGC 7089 to the end of the bridge of NGC 4143 is $\sim$ 1.08 $^{\circ}$, which corresponds to a physical length of $\sim$ 220 kpc. 

The column density map shown in Figure 1 shows that NGC 4111 is at the center of interactions in this group. The higher density HI contours in the center around NGC 4111 indicate that the HI disk direction is inconsistent with that of its stellar disk, implying that it might have experienced strong interactions. The HI gas center for NGC 4111 has deviated from its optical disk as shown in Figure 1. The angular distance between the HI and the center of the optical disk is $\sim$ 1.6$'$ ($\sim$ 7 kpc). Its nearest neighbor galaxies, UGC 7094 and NGC 4117 are the most likely to interact with NGC 4111. To explore this, we show the Position-Velocity plots (P-V diagrams) around NGC 4111 in three directions labeled \textbf{a},\textbf{b}, and \textbf{c} as shown in the upper left panel of Figure 2. Direction \textbf{a} illustrates the relation between NGC 4111 and UGC 7094. The P-V diagram \textbf{a} shows that they have continuous HI distribution in radial velocity direction. They are connected both in the projected plane and in the radial direction, which means that they are interacting together. The P-V diagram \textbf{a} shows that part of the HI gas of NGC 4111 is stripped out of NGC 4111 due to tidal interactions with UGC 7094. The HI gas for UGC 7094 is stretched, too. NGC 4111 has a much lower gas density than UGC 7094 indicated by the intensity level in the P-V plot \textbf{a}. In the east of NGC 4111, we plot the P-V diagram in direction \textbf{b} to investigate the interactions between NGC 4111 and NGC 4117. It shows that their HI gas is linked through a gas bridge in the radial direction, which suggests that NGC 4117 also interacts with NGC 4111. Although no gas component has been removed for NGC 4111 in the direction of NGC 4117, this suggests that their interactions are not as strong as those with UGC 7094. NGC 4111 has a gas density comparable to that of NGC 4117. 
%NGC 4118 does not have continuous velocity distribution with the NGC 4111 system. 

In the south of NGC 4111 and UGC 7094, there is a huge HI gas cloud. There are two major HI components in it, which are marked as cloud A and B. In the upper part of the direction \textbf{c}, we plot the P-V diagram between NGC 4111 and cloud A. It shows that cloud A is also associated with NGC 4111. Part of NGC 4111 HI gas is stripped out to the cloud A direction. Cloud A is the largest HI gas component in the NGC 4111 system. The angular diameter of Cloud A is $\sim$ 12$'$, corresponding to a physical length of $\sim$ 50 kpc. It arises at 730 km$\rm s^{-1}$ next to NGC 4111 and merged with Cloud B after 900 km$\rm s^{-1}$. The highest column density for Cloud A and Cloud B is 4.3$\times 10^{19}\rm cm ^{-2}$ and 3.1$\times 10^{19} \rm cm ^{-2}$, respectively. The region for Cloud A and B spans about $0.5^{\circ} \times 0.5^{\circ}$, where no optical counterpart is found. The estimated HI mass for cloud A + B is 8.44 $\times10^{8} \rm M_{\odot}$. Simulations propose that gas bridges and cloud tails are caused by close encounters \citep{Toomre1972}. Cloud A and B in this system are the much strong observational evidence of tidal interactions between the member galaxies. The lower part of the direction \textbf{c} demonstrates the relationship between cloud A and NGC 4143. NGC 4143 is connected to Cloud A through a HI gas bridge as shown in the P-V diagram \textbf{c} and Figure 1. The HI gas bridge is strong evidence that NGC 4143 has interacted with the NGC 4111 system. The angular length for the bridge is $\sim$36$'$ and corresponds to a projected physical length of 165 kpc. The HI gas density for NGC 4143 is much lower than NGC 4111 as shown in Figure 2c. In this system from NGC 4111 to NGC 4143, the radial velocity for the member galaxies are continuous. 

The galaxy HI mass is derived by the integration of an integral spectrum. In Figure 3 we interpret how we estimate the HI mass for NGC 4111 galaxy from the entire HI gas complex, other galaxies are computed in the same way. The left panel of Figure 3 shows the moment-0 map around the NGC 4111 region. We extract the spectra that are inside the lowest contour as the integral spectrum for NGC 4111, which is shown in the right panel of Figure 3. The central velocity ($V_{\rm sys}$) and velocity width ($w_{\rm 50}$) parameters are obtained by fitting the integral spectra with the Busyfit model \citep{Westmeier2014}. The HI mass for each member galaxy is listed in Table 1. In Table 1 we also list the optical parameters for the member galaxy. We choose 15.1 Mpc as the distance for NGC 4111 galaxy \citep{Roier2022}. For consistency, we use this distance to compute the stellar and HI masses for all the member galaxies as they interact in one group. The stellar masses are estimated using the mass-to-light ratio equation \citep{Bell2003} $\log(M_*/L_r)$=-0.306+1.097(g-r), where $L_r$ is the r band luminosity derived from the absolute magnitude. The HI to stellar mass ratios $f_{\rm HI}$ are multiplied by 100$\%$ for each galaxy.  

\begin{figure*}
    \centering
    \includegraphics[width=0.9\textwidth]{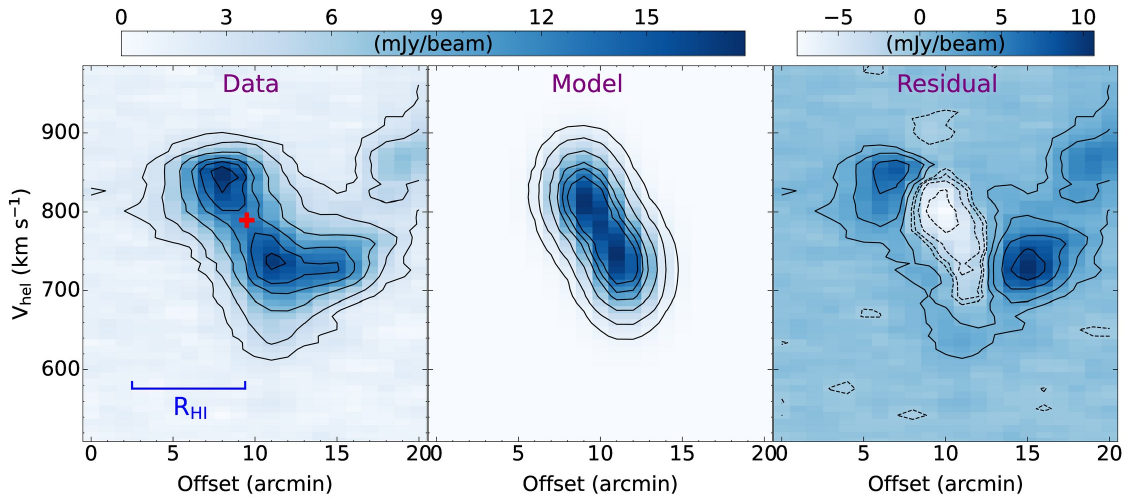}
    \caption{ Comparison between the FAST observational data and the best-fitting model data demonstrated by the HI major axis P-V diagram. Left panel: the P-V diagram extracted from the FAST data cube in the direction of the PA derived from our simulation, the red plus indicates the center of the disk. The radius of the HI disk is labeled. Middle panel: the P-V diagram extracted from the model data cube in the same direction as the left panel. Right panel: the difference between observed and model data which evaluates the performance of our simulation. The contour levels for the left and middle panels are: 0.6, 2.6, 5.4, 8.6, 10.7, 12.9 mJy beam$^{-1}$. The negative contours for the right panel are: -5.4, -2.9, -1.4, -0.7 mJy beam$^{-1}$, and the positive contours are the same as the left and middle panels. }
    \label{fig5}
\end{figure*}

The analyses above elaborate that NGC 4111 has undergone interactions in three directions, which are the UGC 7094, NGC 4117, and Cloud A directions. The HI gas content for NGC 4111 should also be strongly affected. In Figure 4 we display the scaling relations for the member galaxies in this system. In the left panel of Figure 4, we show the relation between stellar mass and HI mass fraction for all member galaxies. The only galaxy whose HI mass is greater than its stellar mass is UGC 7146, which lies outside the NGC 4111 complex. This shows us the power of tidal interactions, that all galaxies in this system have HI masses less than their stellar masses and are HI gas-deficient. The three S0 galaxies are severe HI deficient, whose HI to stellar mass ratio is less than 5\%. The green dashed line is from the empirical relation in nearby group and cluster environments between HI content and stellar mass \citep{Odekon2016}. All the member galaxies in this system lie under this line. In the right panel, we show the \textit{g}-\textit{r} color and stellar mass relation. Here we simply take \textit{g}-\textit{r} = 0.65 as a division of red to blue galaxies \citep{Nelson2018}, as marked by the red dotted line. The color of  NGC 4111 is 0.61 and classified as a blue galaxy, while it is very close to the division line as seen in the right panel of Figure 4. Other S0 galaxies NGC 4117 and NGC 4143 exhibit a red color, and the rest late-type galaxies are in blue.

\section{Discussion}
We have demonstrated that NGC 4111 is currently undergoing strong interactions in this group environment. Here we discuss its effects on the galaxy properties.

First, we focus on the morphology transformation of NGC 4111. The P-V diagram in Figure 5 suggests that it retains some of the rotation structures, implying that it originates from a spiral galaxy. Multi mechanisms have been proposed to explain the formation of S0 galaxies. These mechanisms include ram pressure stripping, mergers, and tidal interactions as stated in the Introduction. NGC 4111 is a member galaxy in this subgroup as well as a member in the Ursa Major cluster. The mass of the Ursa Major is  5$\times10^{13} \rm M_{\odot}$, which is about 1/5 and 1/20 of the Virgo cluster \citep{Trentham2001}.  Research has also suggested that the Ursa Major cluster should be reclassified as a supergroup \citep{Wolfinger2016}. Because the subgroups in Ursa Major are in an early evolutionary state and the majority of the Ursa Major members are HI-rich spiral galaxies. There is not a concentrating core in the Ursa Major, no X-ray emitting intra-cluster gas, and lower velocity dispersion (148 km$\rm s^{-1}$)  are detected \citep{Wolfinger2016}. Thus, mechanisms like ram pressure stripping would have a limited effect on transforming the morphology of NGC 4111. Besides, the polar ring in its inner optical disk of NGC 4111 as mentioned in the Introduction is considered to originate from the minor merger with a gas-rich dwarf galaxy. The gas mass of the dwarf is estimated to be about $10^{7}\rm M_{\odot}$ \citep{Roier2022}. For a gas-rich dwarf galaxy, we assume that the stellar mass is the same order of magnitude as its HI gas mass \citep{Lelli2022}. So the stellar mass for this dwarf galaxy is less than three orders of magnitude of NGC 4111 stellar mass. In general, the minor merger event would have an impact on the morphological type of the major galaxy, while for NGC 4111 it might not be violent enough to change the morphological type from a spiral to an S0 since the mass of the merged dwarf galaxy is so small. We note that there are also two S0 galaxies, NGC 4117 and NGC 4143, in this group, which have no signature of the merger in their optical morphology. These two galaxies are connected with NGC 4111 in the same HI complex, indicating that they might share the same formation mechanism, i.e. tidal interaction within the group. In fact, repeating and continuing tidal interactions with individual nearby galaxies could lead to the destruction of the spiral arms of NGC 4111 and also trigger repetitive star formation and growth of the bulge, leading to transformation to an S0 galaxy. Such a process has been successfully modeled by the work of \citealt{Bekki2011}. This theoretical model also predicts a significant fraction of gas being stripped away from the spiral galaxies and naturally explains the formation of a large amount of intragroup gas we detected in the NGC 4111 group. Similar observational evidence can also be found in the case of NGC 1023 \citep{Xu2023}.  While it is difficult to determine how many interactions NGC 4111  has undergone in the past in this subgroup, the model of \citealt{Bekki2011} suggests that it may take up to 6 Gyr for a spiral to transform to an S0 galaxy. Based on the relatively blue color and its HI gas rotation disk of NGC 4111, we suggest that NGC 4111 is currently in the early transition phase from a spiral to an S0 galaxy despite the possible interaction history. This is consistent with the morphological type of NGC 4111 as SA0(r)$^+$. We propose that tidal interactions in this subgroup might play a more important role than ram pressure stripping and mergers in transforming the morphological type of NGC 4111. Our result provides another observational evidence that tidal interactions in a group environment have promoted the formation of S0 galaxies.

%minor mergers and tidal interactions are considered to play a role in transforming the galaxy morphology.The members are loosely held together and ..  
% The morphological type of NGC 4111 is SA0(r)$^+$, which means that it is currently in the early transition phase from a spiral to an S0 galaxy.
%

%It would be a combined effect that both merger and interactions have led NGC 4111 to start to transform the morphological type. Our result provides another observational evidence that tidal interactions have promoted the formation of S0 galaxies in a group environment.This just provide another argument to support  tidal interaction to be their formation mechanism of the S0 galaxy. 

Second, NGC 4111 has been severely HI-deficient during this transformation process. The $f_{\rm HI}$ for NGC 4111 is only 2.3$\%$ as seen in Table 1. The HI gas for NGC 4111 is almost torn apart due to interactions with UGC 7094, NGC 4117, and Cloud A as stated by the above analysis. Its star formation is suppressed, too; the star formation rate is estimated to be 0.13 $\rm M_{\odot}yr^{-1}$, which is only 5-10\% of that Milky Way \citep{Davis2014}. This result has demonstrated that, in the group environment, the majority of HI gas has been stripped out by interactions with neighboring galaxies when the galaxy morphology starts to transform. The other two S0s in this system NGC 4117 and NGC 4143 also support this point. Their P-V diagrams suggest that these two galaxies still retain their rotation structures. They are also HI deficient; their $f_{\rm HI}$ are 4.0\% and 0.4\%, respectively. NGC 4143 has the lowest HI mass fraction in this system and is missed by previous HI observations. In addition to the member galaxy, the total HI mass in the whole NGC 4111 system (except UGC 7146, including cloud A and B and diffuse gas) is 3.0$\pm0.4\times10^{9}\rm M_{\odot}$ and the corresponding total stellar mass is 4.99$\times10^{10} \rm M_{\odot}$. The $f_{\rm HI}$ is 6.0$\%$ for the whole system, which is also a very low HI gas fraction. This implies that most of the HI gas for each galaxy has been stripped out of this system. The stripped gas either exists as tidal clouds in the intergalactic medium (IGM) or is ionized by UV photons in outer space and becomes too diffuse to detect.
%becomes too diffuse to detect distribution in space. 

Third, we created a 3D model for NGC 4111 using the 3D tilted ring fitting by TiRiFiC software \citep{Jozsa2007} to investigate the HI kinematical features in NGC 4111. This software can model our cube data through a series of concentric rings, described by a comprehensive set of geometrical parameters (kinematical center, inclination \textit{i}, and position angle PA) and kinematic parameters (systemic velocity $V_{\rm sys}$, velocity dispersion $V_\sigma$ and rotation velocity $V_{\rm rot}$). During the simulation, we fixed the kinematic center and the system velocity based on the velocity field. The other parameters are set free. The comparison between our observations and the best-fitting model is shown in Figure 5. The HI disk position angle for NGC 4111 is simulated as 11.0$\pm$5.3${^ \circ}$, inconsistent with the stellar disk of 151.4$^{\circ}$. This indicates that the HI gas disk does not rotate with its optical disk. The optical disk and HI major axis direction for NGC 4111 are marked as a blue ellipse shape and a blue dashed line respectively in the left panel of Figure 3. The direction of the HI gas rotates differs from its stellar disk about 34.2$^{\circ}$ for NGC 4111. In the left panel of Figure 5, we display the P-V diagram in the simulated HI major axis (11$^{\circ}$) from the FAST data cube, which shows a good rotation disk structure. The rotational structure implies that it originates from a spiral galaxy. The radius of the HI disk is about 7$'$ which is labeled in the left panel of Figure 5, corresponding to a physical radius of about 30.7 kpc. In the middle panel, we use the best-fitting model data cube to plot the major axis P-V diagram in the same direction. The right panel is the difference between the data and the best-fitting model. Because NGC 4111 is currently undergoing a transition from a spiral galaxy to an S0, its rotation structure is destructed by tidal interactions. So the main source of the residual comes from the disk rotation deviating from a standard spiral disk rotation structure. Nevertheless, we can still discern the rotational structure of the galaxy disk. Besides that, the residual is relatively small suggesting that our simulation is reasonable.  

%It is noteworthy that the HI to stellar mass ratio $f_{\rm HI}$ is extremely low for the SO galaxies in this system, NGC 4111, NGC 4117 and NGC 4143. The velocity width $w_{50}$ of NGC 4111 is fitted to be 180 km $\rm s^{-1}$ as seen in Table 1.

Furthermore, we estimate the dynamical mass for NGC 4111. In our simulation, the rotational velocity and velocity dispersion for NGC 4111 HI disk are simulated to be 56.1$\pm$4.9 and 37.2$\pm$3.6 km $\rm s^{-1}$, respectively. The radius ($R_{\rm HI}$) of the HI disk is chosen as 30.7 kpc estimated from Figure 5. The dynamical mass for NGC 4111 is derived to be $5.8\times10^{10}\rm M_{\odot}$ using the formula $M_{\rm dyn} = (V_{\rm rot}^2 + 3\sigma_V^2)R_{\rm HI}/G$, where $V_{\rm rot}$ is the rotation velocity and $\sigma_V$ is velocity dispersion. We check the simulated rotation velocity by the 
empirical relation between rotation velocity and velocity width of HI gas. The velocity width $w_{50}$ has been widely used as an indicator of the rotation velocity (\citealt{Tully1997}; \citealt{Begeman1989};\citealt{Broeils1997}; \citealt{Spekkens2018}) and is used for estimating the dynamical mass of galaxies. The rotation velocity is calculated as $V_{\rm rot} = \frac{w_{\rm 50}}{2\times \rm sin(\textit{i})} $, where \textit{i} is the inclination of the galaxy. Thus the dynamical mass for NGC 4111 based on velocity width is $5.2\times10^{10}\rm~M_{\odot}$, which is roughly equal to that derived from the simulation, suggesting that the simulated rotation velocity is reasonable.  The ratio of rotation velocity and velocity dispersion ($V_{\rm rot}/\sigma$) is $\sim$1.5 for NGC 4111. It is much lower than the Milky Way-like spiral galaxies, which are in the range of 10-20 \citep{Epinat2010}. It implies that NGC 4111 is experiencing a transition from a rotational supported to a pressure-supported early-type galaxy (\citealt{Genzel2011}; \citealt{Johnson2018}).

%We use $w_{50}$ instead of $w_{20}$ to estimate the dynamical mass because in the case of NGC 4111 $w_{20}$ would be untruly wide caused by the interactions of its neighbors.

After subtracting the stellar and HI gas mass the dark matter mass is 3.9$\pm$0.5$\times10^{10}\rm M_{\odot}$. Then the dark matter-to-stellar mass ratio (r$_{\rm DM}$) is computed as 3.1$\pm$0.7, implying a small ratio. In the local Universe the average r$_{\rm DM}$ is about 20-30 \citep{Moster2010} for the galaxies whose stellar masses are similar to that of the Milky Way. Dark matter deficient galaxy is a trending topic since NGC 1052 DF2 is found \citep{van Dokkum2018}. NGC 1052 DF2 is a low-brightness galaxy in the NGC 1052 group, which is almost absent of dark matter. The r$_{\rm DM}$ for NGC 1052 DF2 is of order unity. Plenty of mechanisms have been proposed to explain the abnormal features of NGC 1052 DF2. Among them, the most compelling explanation is that the dark matter was stripped during the interactions with its neighbor galaxies. NGC 4111 is also a dark matter-deficient galaxy, whose r$_{\rm DM}$ is much lower than the average pointed out by cosmological simulations. For NGC 4111 which lies in this group environment, we speculate the most possible reason would be that its dark matter was stripped out by interactions. This is the first time that we find both dark matter and HI gas being stripped simultaneously by tidal interactions in the meantime the morphology transforms from spiral to S0. 

%It is surprised to find that galaxies interactions in group environment have significant effect on the galaxy properties. Both the HI gas and dark matter for NGC 4111 has been stripped out in the meantime galaxy morphology type transforms from spiral to S0.   

%SDSS J120625.42+422607.2 is a dark galaxy that first detected by HI, labeled in Figure 1. There is no distinct bridge between Cloud D and SDSS J120625.42+422607.2 detected like that for NGC 4143. While there are several weak and diffuse gas clouds between them. The HI mass fraction for SDSS J120625.42+422607.2 is also very low, only 3.54$\%$. It is the smallest and faintest galaxy in this system. 

%However, the rest spiral and dwarf galaxies have relatively higher HI gas fraction than the S0 galaxies. It is the bluest galaxy in this system while it is the first time detected by HI. The HI map indicates that it might have interacted with Cloud D, and most of its HI gas are stripped during this process.

%(the sum of eight galaxies HI mass) is 2.86$\times10^{9} 

%Even though this ratio is not as low as the famous ultra diffuse dark matter deficient galaxy NGC 1052-DF2 whose ratio is of order unity, the r$_{dm}$ for NGC 4111 is really a small ratio. 

%Simulating their relative motion of each galaxy is challenging because too many galaxies are involved. The most probable picture is like this. 

%\section{Conclusion}

\section{acknowledgments} 
We thank the FAST staff for help with the FAST observations. We acknowledge the support of the China National Key Program for Science and Technology Research and Development of China (2022YFA1602901). This work is  supported by the National Natural Science Foundation of China (Grant Nos. 12373001, 12225303, 12421003 and 12033004), the Chinese Academy of Sciences Project for Young Scientists in Basic Research, grant no. YSBR-063, the Youth Innovation Promotion Association of CAS, and the Central Government Funds for Local Scientific and Technological Development (No. XZ202201YD0020C).  This work is supported by the Guizhou Provincial Science and Technology Projects (QKHFQ[2023]003,QKHPTRC-ZDSYS[2023]003,QKHFQ[2024]001-1). We also acknowledge support from the National Key $R\&D$ Program of China (2018YFE0202900; 2017YFA0402600).  This work was supported by the Open Project Program of the Key Laboratory of FAST, NAOC, Chinese Academy of Sciences. %%%%%%%%%%%%%%%%%%%%%%%%%%%%%%%%%%%%%%%%%%%%%%%%%%



\bibliographystyle{apj}

%%%%%%%%%%%%%%%%%%%%%%%%%%%%%%%%%%%%%%%%%%%%%%%%%%

%%%%%%%%%%%%%%%%% APPENDICES %%%%%%%%%%%%%%%%%%%%%
%\newpage

%%%%%%%%%%%%%%%%%%%%%%%%%%%%%%%%%%%%%%%%%%%%%%%%%%

% Don't change these lines
\label{lastpage}
\end{document}